\documentclass{phostproc}
\usepackage{hyperref}
\usepackage{url}
\graphicspath{{figures/}}

\title{Impact of general differential rotation on gravity waves in rapidly rotating stars}
\author{Vincent Prat,$^{1}$ St\'ephane Mathis,$^{1}$ Kyle Augustson,$^{1}$ Fran\c cois Ligni\`eres,$^{2,3}$ J\'er\^ome Ballot,$^{2,3}$ Lucie Alvan,$^{1}$ Allan Sacha Brun$^{1}$}

\affiliation{$^{1}$ AIM, CEA, CNRS, Universit\'e Paris-Saclay, Universit\'e Paris Diderot, Sorbonne Paris Cit\'e, F-91191 Gif-sur-Yvette, France \\
			 $^{2}$ Université de Toulouse, UPS-OMP, IRAP, Toulouse, France \\
             $^{3}$ CNRS, IRAP, 14 avenue \'Edouard Belin, 31400 Toulouse, France}

\shortauthors{Vincent Prat \textit{et al.}} %to be used for contributions with 3 authors and more.

\abs{Differential rotation plays a key role in stellar evolution by triggering hydrodynamical instabilities and large-scale motions that induce transport of chemicals and angular momentum and by modifying the propagation and the frequency spectrum of gravito-inertial waves.
It is thus crucial to investigate its effect on the propagation of gravity waves to build reliable seismic diagnostic tools, especially for fast rotating stars, where perturbative treatments of rotation fail.
Generalising a previous work done in the case of uniform rotation, we derived a local dispersion relation for gravity waves in a differentially rotating star, taking the full effect of rotation (both Coriolis and centrifugal accelerations) into account.
Then we modelled the propagation of axisymmetric waves as the propagation of rays.
This allowed us to efficiently probe the properties of the waves in various regimes of differential rotation.
}

\begin{document}

\maketitle

\section{Introduction}

Internal gravity waves are a unique way to probe stellar interiors for intermediate-mass and massive stars.
Indeed, they provide us with constraints on mixing and rotation thanks to frequencies of gravity modes \citep[e.g.][]{VanReeth18}.
They are also able to transport angular momentum in all stellar types (see \citealt{Schatzman,Zahn97,TalonCharbonnel05} for late-type stars, \citealt{Lee,Rogers13,Rogers15} for early-type stars, and \citealt{TalonCharbonnel08, Fuller, Pincon} for evolved stars).
Finally, tidally excited internal gravity waves contribute to tidal dissipation in close-in binary and planetary systems \citep[e.g.][]{Zahn75, OgilvieLin04, OgilvieLin07}.

Differential rotation affects the dynamics of stellar interiors in several ways.
First, it modifies the propagation of gravity waves, their frequencies and the transport they induce \citep[e.g.][]{Ando, LeeSaio93, Mathis09, Mirouh, Guenel}.
Second, it triggers hydrodynamical instabilities that contribute to the transport of angular momentum and chemical elements in stellar radiative zones \citep[e.g.][]{Zahn92, MaederZahn, MathisZahn}.
For those reasons, it is crucial to have a reliable description of gravity waves in presence of differential rotation to be able to constrain transport processes and rotation in differentially rotating stars.

For slow rotators, the effect of rotation can be considered as a perturbation of the non-rotating system, and the theory predicts the splitting of modes of same radial order and angular degree but different azimuthal orders.
This allowed us to estimate the near-core rotation rate of a large number of subgiant and red giant stars thanks to mixed modes \citep{Beck,Mosser,Deheuvels12, Deheuvels14, Deheuvels15, Triana17, Gehan}.
For some massive stars, constraints on the internal rotation rate have been obtained thanks to splittings of gravity and sometimes pressure modes \citep{Kurtz, Saio15, Triana15, Murphy}.

For fast rotators, however, pertubative methods are no longer valid \citep[see][]{Reese06,Ballot10, Ballot13}.
The computation of modes then requires solving a full two-dimensional (2D) problem, which is numerically more expensive.
This concerns typical intermediate-mass and massive pulsators, such as $\gamma$ Doradus, $\delta$ Scuti, slowly pulsating B (SPB), $\beta$ Cephei, and Be stars.
Thus, new seismic diagnoses need to be built for rapidly rotating stars \citep[e.g.][]{Ouazzani17}.

To go beyond pertubative methods without solving the full 2D problem, one may use the traditional approximation of rotation (TAR), which neglects the radial component of the Coriolis acceleration.
Doing so makes the 2D problem separable again in spherical coordinates, thus allowing for efficient computation of modes in the case of uniform rotation \citep{LeeSaio97, Townsend03, Bouabid}.
This approximation has been used to interpret seismic data of $\gamma$ Doradus stars \citep{VanReeth16} and SPB stars \citep{Papics17}.
\citet{Mathis09} proposed a way to include differential rotation in the formalism of the TAR, and \citet{VanReeth18} applied it to $\gamma$ Doradus stars.
However, the domain of validity of this approximation is still unclear.

Another approach is the ray dynamics, which models small-wavelength propagating waves as ray trajectories, similarly to geometrical optics.
This approach has been used first for acoustic waves in rapidly, uniformly rotating stars by \citet{LG09} and \citet{Pasek12}.
Later, it has been used for gravity waves by \citet{PratLB} in the uniformly rotating case.
This study predicted the existence of three families of modes: (i) regular modes, which are similar to modes in non-rotating stars; (ii) island modes, where the energy is localised around periodic orbits; and (iii) chaotic modes, which have irregular spatial patterns.
Each family is expected to have its own spectral regularities, and period spacings have been derived for low-frequency regular modes \citep{PratMLBC}.

We present here the generalisation of this work to the case of different rotation, published in \citet{PratMALBAB}.
In Sect.~\ref{sec:method}, we explain the ray theory.
In Sect.~\ref{sec:results}, we show the results.
Finally, we conclude in Sect.~\ref{sec:conclusion}.

\section{Ray theory}
\label{sec:method}

We use a ray theory based on the Hamiltonian dynamics.
Given a local dispersion relation $\omega=\omega(\vec x, \vec k)$, where $\omega$ is the angular frequency of a wave, $\vec x$ is the position, and $\vec k$ the wavevector, the ray trajectory follows the group velocity of the wave:
\begin{equation}
    \frac{ {\rm d}\vec x}{ {\rm d}t} = \vec\nabla_{\!\vec k}\omega
\end{equation}
and the wavector evolves along the ray path so that the angular frequency is conserved:
\begin{equation}
    \frac{ {\rm d}\vec k}{ {\rm d}t} = -\vec\nabla_{\!\vec x}\omega.
\end{equation}
To visualise the structure of the phase space $(\vec x,\vec k)$, we use Poincaré surfaces of section (PSS) that are obtained by taking the intersection of ray trajectories with a given surface (in our case the equatorial plane $\theta=\pi/2$, where $\theta$ is the colatitude) at a given frequency.
For axisymmetric waves, the dynamics is 2D, and the PSS can be represented in the $(r, k_r)$ plane, where $r$ is the spherical radial coordinate and $k_r$ the radial component of the wavevector.

\begin{figure}
    \centering
    \includegraphics[width=0.75\linewidth]{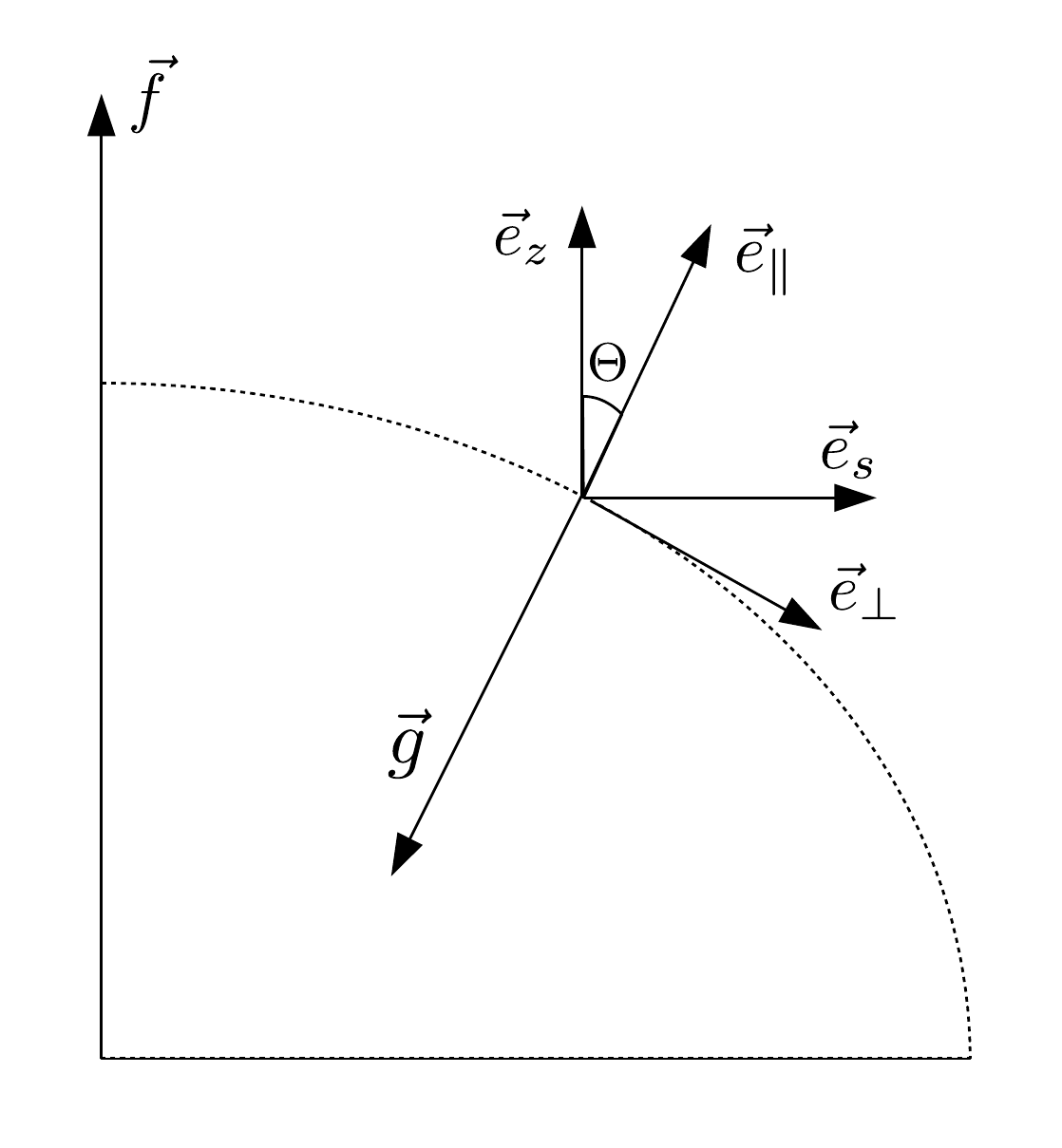}
    \caption{Illustration of the coordinate systems}
    \label{fig:coord}
\end{figure}

In \citet{PratMALBAB}, we derived a general dispersion relation including the full Coriolis acceleration, differential rotation, centrifugal deformation, stable stratification, and surface effects responsible for the back-refraction of waves:
\begin{eqnarray}
    (k^2+k_{\rm c}^2)\omega^2\hskip-0.9em &= f(f+Q_s)k_z^2-fQ_z(k_sk_z+k_\parallel k_\perp)    \nonumber\\
                                &   +{N_0}^2k_\perp^2+f\cos\Theta(f\cos\Theta+Q_\perp)k_{\rm c}^2,
\end{eqnarray}
where $k$ is the norm of the wavevector, $k_z$ and $k_s$ its components parallel and orthogonal to the rotation axis, $k_\parallel$ and $k_\perp$ its components parallel and orthogonal to the effective gravity (which includes the effect of the centrifugal acceleration), $f=2\Omega$ the Coriolis frequency, $\Omega$ the rotation rate, $\vec Q=r\sin\theta\vec\nabla\Omega$, $N_0$ the Brunt-V\"ais\"al\"a frequency, $\Theta$ the angle between the rotation axis and the direction of the gravity vector, $k_{\rm c}$ the surface term defined by
\begin{equation}
    k_{\rm c}^2 = \frac{\Gamma_1^2}{4} \frac{\mu-1}{\mu+1} \frac{g_0^2}{c_{\rm s}^4},
\end{equation}
$\Gamma_1$ the first adiabatic exponent, $\mu$ the polytropic index, $g_0$ the norm of the effective gravity vector, and $c_{\rm s}$ the sound speed.
The coordinate system is illustrated in Fig.~\ref{fig:coord}.

\section{Results}
\label{sec:results}

In this section, we present results obtained by performing ray-tracing computations in rotating polytropic models, which are characterised by a relationship between the pressure $P$ and the density $\rho$ of the form $P\propto\rho^{1+1/\mu}$.
For simplicity reasons, we neglected the effect of differential rotation on the background models and used arbitrary rotation laws.
We investigated first the effect of a purely radial differential rotation (Sect.~\ref{sec:radial}) and second that of latitudinal differential rotation as well (Sect.~\ref{sec:latitudinal}).

\subsection{Radial differential rotation}
\label{sec:radial}

\begin{figure*}[ht]
	\centering
	\includegraphics[width=\linewidth]{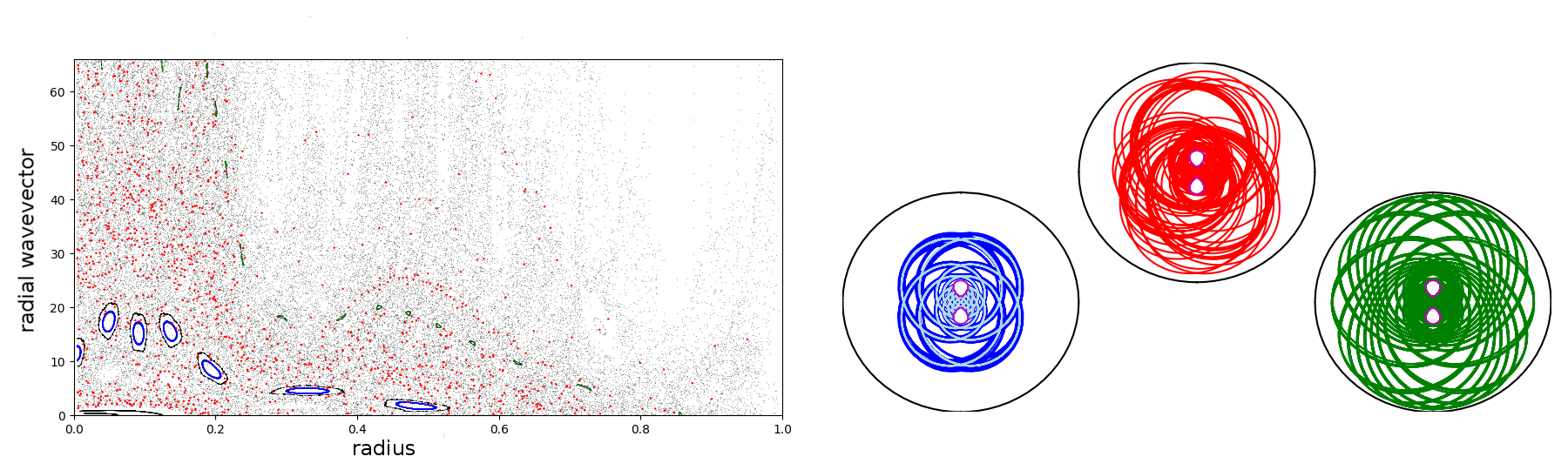}
    \caption{PSS (left) and trajectories (right) for a stellar model rotating at 38\% of its critical velocity, with $\Omega_{\rm C}/\Omega_{\rm R}=2$ and $\omega/f=1.6$.}
	\label{fig:fast}
\end{figure*}

\begin{figure*}[ht]
	\centering
	\includegraphics[width=\linewidth]{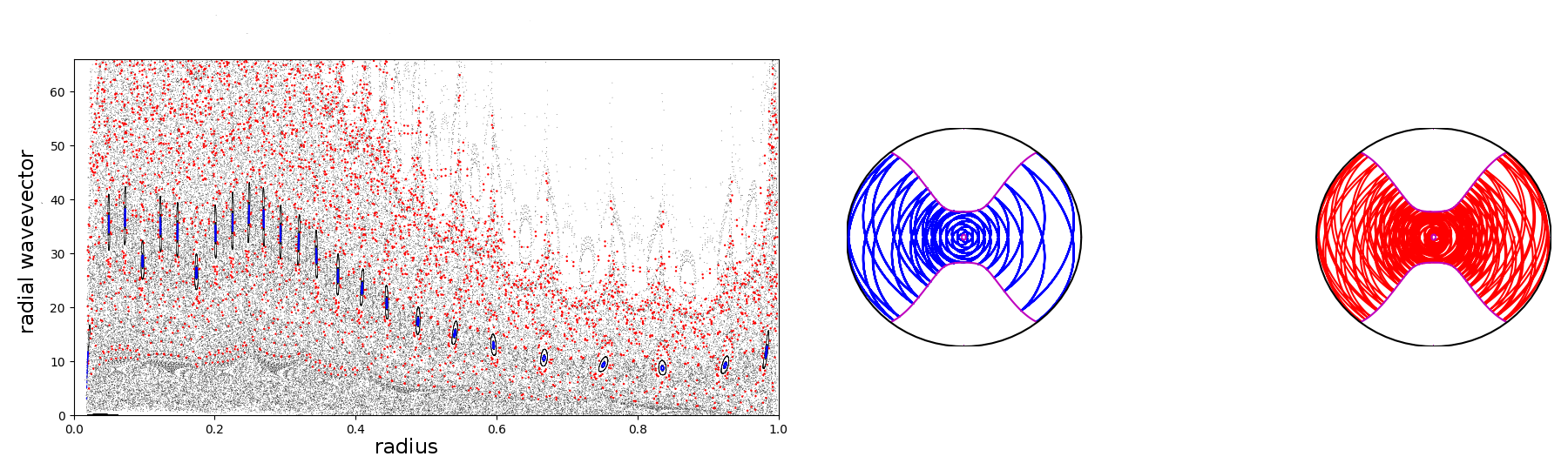}
    \caption{Same as Fig.~\ref{fig:fast}, but with $\Omega_{\rm C}/\Omega_{\rm R}=0.5$ and $\omega/f=0.8$.}
	\label{fig:slow}
\end{figure*}

We tested two configurations: one with a core rotating faster (at a rate $\Omega_{\rm C}$) than the rest of the star (at a rate $\Omega_{\rm R}$), and one with a slower core.
In both cases, when the wave frequency is larger than the maximum of the Coriolis frequency, we observed that the wave dynamics is similar to the uniformly rotating case in the super-inertial regime.
Likewise, when the wave frequency is smaller than the minimum of the Coriolis frequency, the wave dynamics recalls that of the sub-inertial regime in the uniformly rotating case.
The novelty of the differentially rotating case appears when considering wave frequencies that are between the minimum and the maximum of the Coriolis frequency.
This new regime, which we call trans-inertial, is characterised by a large fraction of chaotic trajectories and a few island chains, as illustrated in Figs.~\ref{fig:fast} and \ref{fig:slow} for the fast and slow cores, respectively.
These figures also show new kinds of propagation domains found in the trans-inertial regime.
Although the propagation domains are very different in the two cases, the dynamics looks similar.

\subsection{Latitudinal differential rotation}
\label{sec:latitudinal}

We consider here a rotation profile with a uniformly rotating inner region (at the rate $\Omega_{\rm R}$) and an outer region with a latitudinal differential rotation defined by $\Omega = \Omega_{\rm R} + \Omega_{\rm D}\cos(2\theta)$.
This rotation profile generates many different regimes that depend on the wave frequency and the degree of differential rotation $\Omega_{\rm D}/\Omega_{\rm R}$.
We will focus here on trans-inertial regimes, since the purely sub-inertial and super-inertial regimes are similar to the corresponding regimes in the uniformly-rotating case.

\begin{figure*}[ht]
    \centering
    \includegraphics[width=\linewidth]{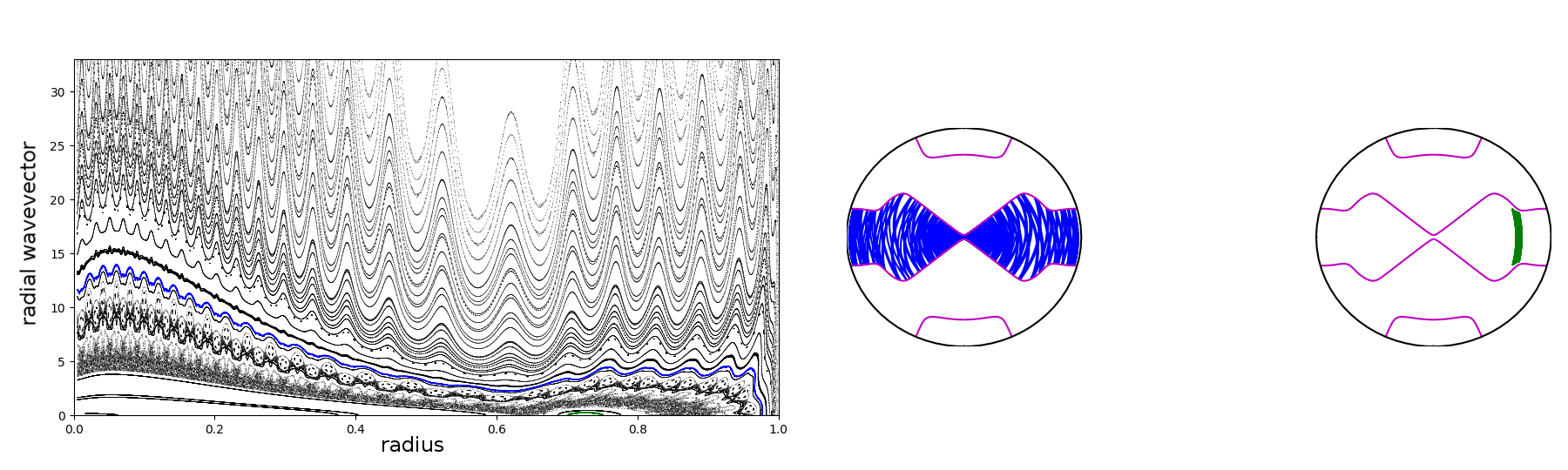}
    \caption{PSS (left) and trajectories (right) for a stellar model rotating at 38\% of its critical velocity, with $\Omega_{\rm D}/\Omega_{\rm R}=-0.74$ and $\omega/f=0.61$.}
    \label{fig:pseudosub}
\end{figure*}

One of the trans-inertial regimes leads to three disconnected propagation domains: one near each pole and one near the equatorial plane.
The dynamics of rays computed in the equatorial region is mostly regular, similarly to that of the sub-inertial regime, although the PSS (shown in Fig.~\ref{fig:pseudosub}) has many fine structures and some chaos.

\begin{figure*}[ht]
    \centering
    \includegraphics[width=\linewidth]{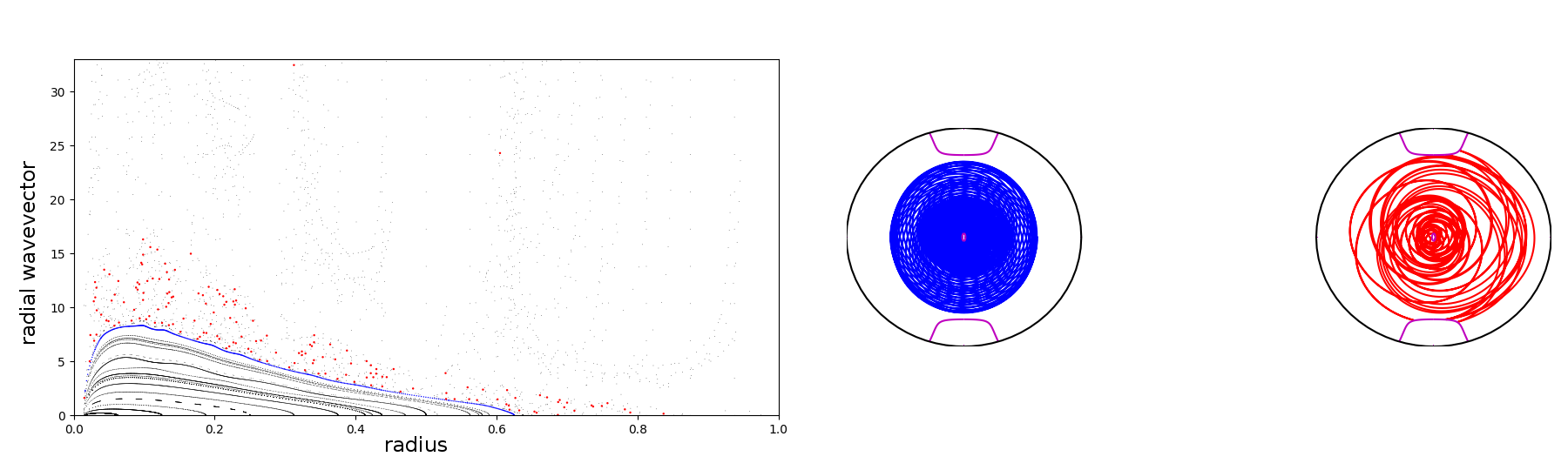}
    \caption{Same as Fig.~\ref{fig:pseudosub}, but with $\Omega_{\rm D}/\Omega_{\rm R}=0.22$ and $\omega/f=1.11$.}
    \label{fig:trans_lat}
\end{figure*}

Other trans-inertial regimes have various propagation domains, but similar dynamics, illustrated in Fig.~\ref{fig:trans_lat}: regular trajectories at low $k_r$ when rays do not propagate in the differentially rotating region, and chaotic ones at larger $k_r$ when rays do propagate in the differentially rotating region.

\section{Conclusion}
\label{sec:conclusion}

The general dispersion relation we derived allowed us to efficiently explore the phase space in a large parameter range.
This exploration showed that purely sub- and super-inertial regimes are similar to their uniformly rotating equivalents.
In contrast, trans-inertial regimes, which exist only in the differentially rotating case, show very different propagation domains, and a dynamics dominated by chaotic trajectories.

The large variety of propagation domains means that differential rotation probably has a significant impact on the mode frequencies, and that this impact might be observable in real oscillation spectra.
For similar reasons, differential rotation may have an impact on the excitation and damping of waves, and thus on the mode amplitudes, on the transport of angular momentum and on the tidal dissipation by gravito-inertial waves.

The fact that the low-frequency dynamics (the sub-inertial regime) is dominated by regular trajectories suggests that it might be possible to derive new semi-analytical diagnoses for differential rotation as it has been done for uniform rotation \citep{PratMLBC}.
Such diagnoses could be compared to the work of \citet{VanReeth18}, who used the traditional approximation of rotation.

One limitation of the present work is that the considered stellar models are not realistic.
First, the effect of differential rotation on the background model has been neglected, and the rotation profile is not coupled to the structure.
Second, we considered fully radiative models.
One way to go further would be to use more realistic rotating models such as those produced by the ESTER code \citep{Rieutord16}, in which structure and rotation are inherently coupled in two dimensions.

We presented here axisymmetric waves, but \citet{PratMALBAB} also derived a dispersion relation for non-axisymmetric waves.
The dynamics of those waves still is to be explored, and this is a crucial step towards a description of the transport of angular momentum by gravito-inertial waves based on the ray theory, since the transport arises from the different damping of prograde and retrograde waves.

\section*{Acknowledgments}
V.P., S.M., and K.A. acknowledge support from the European Research Council through ERC grant SPIRE 647383.
V.P., F.L., and J.B. acknowledge the International Space Science Institute (ISSI) for supporting the SoFAR international team\footnote{\url{http://www.issi.unibe.ch/teams/sofar/}}.
The authors acknowledge funding by SpaceInn, PNPS (CNRS/INSU), and CNES CoRoT/Kepler and PLATO grants at DAp and IRAP.

\bibliographystyle{phostproc}
\bibliography{biblio}

\end{document}